\documentclass[12pt,a4paper,dvips]{article}
\usepackage{a4p}
\usepackage{cite,mcite}
\usepackage{graphicx}
\usepackage{physics}
\usepackage{amsmath}
\usepackage{dcolumn}
\usepackage{l3_title,ifthen,Lep}
\journalname{Phys. Lett. B} 

\date{October 31, 2001}
\preprint{2001-075}
\Lep{2}
\newlength{\capindent}
\newlength{\capwidth}
\newlength{\figwidth}
\newcommand{\icaption}[2][!*!,!]{\hspace*{\capindent}%
  \begin{minipage}{\capwidth}
    \ifthenelse{\equal{#1}{!*!,!}}%
      {\caption{#2}}%
      {\caption[#1]{#2}}
  \end{minipage}}
\newcommand{\GG}{\gamma^*\gamma^*}
\newcommand{\GGG}{\gamma\gamma}

\newcommand{\sgg}{\sigma_{\GG}}
\def\EE{\ifmath{\mathrm{e^+ e^-}}}
\def\TT{\ifmath{\mathrm{\tau^+ \tau^-}}}%

 {\end{list}}
\begin{document}
\begin{titlepage}
\title{Double-Tag Events in Two-Photon Collisions at LEP}
\author{The L3 Collaboration}
%
%
\begin{abstract}
 
  Double-tag events in two-photon collisions are studied using the L3 detector
  at LEP centre-of-mass energies from $\sqrt{s}=189 \GeV$
  to 209~GeV. The cross sections of the
  $\EE \rightarrow \EE hadrons$ and $\GG \rightarrow hadrons$ processes 
  are measured as a function of the product of the photon virtualities,
  $Q^2 = \sqrt{Q^2_1Q^2_2}$,
  of the two-photon mass, $W_{\GGG}$, and of the variable
  $Y= \ln (  W^2_{\GGG} /{Q^2 })$. The average photon virtuality
  is $\langle Q^2_1 \rangle = \langle Q^2_2 \rangle = 16 \GeV^2$. 
  The results are in agreement with  
  next-to-leading order calculations for the process 
  $\gamma^* \gamma^* \ra \mathrm{q}\qbar$ in the interval $2 \le Y \le 5$.  
  An excess is observed in the interval $5 < Y \le 7$, corresponding to 
  $W_{\GGG}$ greater than 40 \GeV . This may be interpreted as a contribution
  of resolved photon QCD processes or the onset of BFKL phenomena.

\end{abstract}
%
%
\submitted

\end{titlepage}
%
%
\section{Introduction}

  This letter presents a measurement of cross sections of two-photon 
  collisions:~$\EE \rightarrow \EE hadrons$ obtained with the L3 
  detector~\cite{l3det} using
  double-tag events, where both scattered electrons
  \footnote{Electron stands for electron or positron throughout this paper.}
  are detected in the small angle electromagnetic calorimeters.
  The virtualities of the two photons are defined as 
  $Q^2_i = 2E_iE_b(1-\cos\theta_i)$, where $E_{b}$ is the beam energy,
  $E_i$ and $\theta_i$ are the measured energy and scattering angle of the 
  detected electron ($i=1$) or positron ($i=2$).  The centre-of-mass energy of 
  the two virtual photons, $W_{\GGG}$, is related to the $\EE$ centre-of-mass 
  energy, $\sqrt{s}$, by $W^2_{\GGG} \approx s y_1y_2$, with 
  $y_i = 1-(E_i/E_{b})\cos^2(\theta_i/2)$. This is a good approximation in the
  kinematic range covered by this study, where $W^2_{\GGG}$ is usually much 
  larger than $Q^2_i$.
  It is convenient to define the dimensionless variable $Y$:    
  \begin{equation}
    Y = \ln{\frac{W^2_{\GGG} }{Q^2}}, ~~~ Q^2=\sqrt{Q^2_1 Q^2_2}
  \end{equation} 
  which depends mainly on the angles of the scattered electrons 
  and allows the combination of the 
  data collected at different values of $\sqrt{s}$.

  Taking advantage of the good energy resolution of the small angle
  electromagnetic calorimeters, $W_{\GGG}$ is 
  calculated as the missing mass of the two scattered electrons, 
  $W_{\mathrm{ee}}$. This avoids an unfolding procedure to calculate
  $W_{\GGG}$ from the effective mass of the hadrons observed in the detector,
  $W_{vis}$, which is the dominant source of systematic uncertainty on the 
  measurement of the $\EE \rightarrow \EE hadrons$ cross sections for 
  untagged~\cite{l3tot,opaltot} and single-tag~\cite{L3F2,LEPF2} events. 
  However the $W_{\mathrm{ee}}$ variable is
  more strongly affected by QED radiative corrections than $W_{vis}$. 

\begin{figure}[htbp]

\vfil
   \begin{center}
     \includegraphics[height=0.75\textheight]{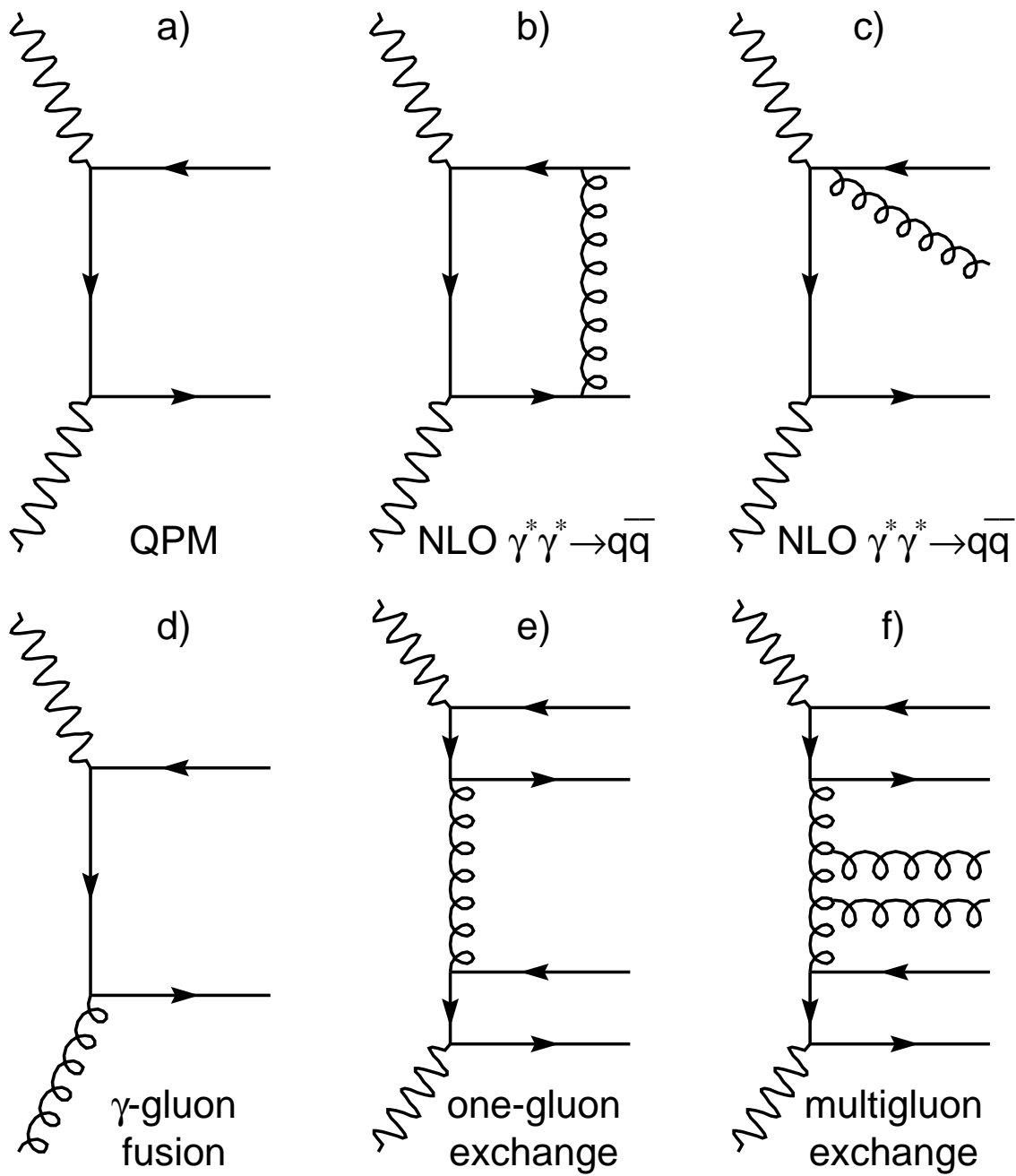}
   \end{center}
 
  \caption[]{Examples of diagrams contributing to the process $\gamma ^* \gamma^* \ra hadrons $ :
             a) QPM, b) and c) $\mathcal{O}(\as)$ QCD corrections to the QPM diagram,
             d) photon-gluon fusion,  
             e) one-gluon exchange and
             f) multigluon ladder exchange. 
  \label{fig:diagram} }

\end{figure}

  In perturbation theory, the cross section of the 
  $\gamma^* \gamma^* \ra   hadrons$
  process is described in terms of a fixed order expansion in the strong
  coupling constant, complemented with 
  the DGLAP~\cite{dglap} evolution of the parton density of the photon.
  All two-to-two leading order (LO) processes, such as
  $\gamma \gamma \ra  \mathrm{q}\qbar $ (QPM) or,
  for example, $\gamma g \ra \mathrm{q} \qbar$ or
  $\gamma \mathrm{q} \ra g \mathrm{q}$
  (single resolved) and $g g \ra \mathrm{q} \qbar $ (double resolved),
  are taken into account in the Monte Carlo
  generators used to analyse the data. If the virtualities of the two 
  photons are large and comparable, LO processes are expected to be 
  suppressed relative to diagrams where multiple gluons are exchanged between 
  the $\mathrm{q} \qbar$ dipoles~\cite{dipole} coupling to each virtual 
  photon. Examples of possible diagrams are given in Figure~\ref{fig:diagram}.
  In leading logarithmic approximation, the resummed series of
  perturbative gluonic ladders can be described 
  by the BFKL equation~\cite{bfkl}, which predicts a rise in cross sections 
  as a power of $W_{\GGG}$, as if a ``hard Pomeron''~\cite{hard} was 
  exchanged. The cross section measurement of two virtual photons is 
  considered as a ``golden'' process to test BFKL dynamics~\cite{gg2}.
  After our first publication on the double-tag data at
  $\sqrt{s}=91\GeV$ and 183~GeV~\cite{paper_168}, an effort was made
  to improve the QPM calculation by including QCD corrections~\cite{cacciari}.
  The effects of varying the charm mass and the strong coupling constant 
  were studied as well as the contribution of longitudinal 
  photon polarization states~\cite{lo}. The relative importance of 
  perturbative and non perturbative QCD effects was also addressed by 
  considering Reggeon and
  Pomeron contributions~\cite{donn,kw}. There are also many efforts to include 
  next-to-leading order (NLO) corrections in the BFKL model~\cite{nlo}.
  
  The data, discussed in this letter, were collected at
  $\sqrt{s} = 189 - 209\GeV$ and correspond to an integrated 
  luminosity of 617~pb$^{-1}$, for a luminosity weighted centre-of-mass energy
  $197.9 \GeV$. The observed value of $Q^2_i$ is in the range 
  $4 - 44 \GeV^2$ with an average value of
  $\langle Q^2_i \rangle = 16 \GeV^2$. The kinematic region 
  $E_{1,2} > 40 \GeV$, $30~\mathrm{mrad} < \theta_{1,2} < 66~\mathrm{mrad}$
  and $W_{\GGG} > 5 \GeV$ is investigated.
  A study of asymmetric double-tag events ($Q^2_1 \gg Q^2_2$) 
  at $\sqrt{s}=91\GeV$ was previously reported~\cite{paper_169}.

\section{Event Generators}

  Two Monte Carlo generators, PHOJET~\cite{pho} and TWOGAM~\cite{two}, 
  are used to simulate double-tag two-photon events. Both use the 
  GRV-LO~\cite{grv} parton density in the photon
  and include all two-to-two LO $\gamma \gamma$ diagrams. They   
  describe well single-tag events~\cite{L3F2}. 

  PHOJET is an event generator for pp, $\gamma$p and  two-photon interactions, 
  based on the Dual Parton Model. 
  The electron-photon vertex for transversely polarized photons~\cite{budnev}
  is simulated. A transverse momentum cutoff of
  2.5~GeV on the outgoing partons is applied to 
  separate soft from hard processes~\cite{FKP}. 
  PHOJET gives also a good description of untagged
  $\gamma\gamma \rightarrow hadrons$ events~\cite{l3tot}.
  The electromagnetic coupling constant, $\alpha_{em}$, in PHOJET is
  fixed to the value for on-shell photons.
  
  \par
  TWOGAM generates three different processes separately: QPM,
  QCD resolved photon processes and non perturbative soft processes described
  by the Vector Dominance Model (VDM).
  The normalization of the QPM process is determined by the quark masses
  ($m_u=m_d = 0.3 \GeV$, $m_s=0.5 \GeV$ and $m_c=1.6\GeV$), that of the VDM 
  process is fixed by our measurement of the cross section of real 
  photons~\cite{l3tot}, while the normalization of the QCD contribution is
  adjusted to reproduce the observed number of data events.
  TWOGAM was recently upgraded to take into account QED soft and hard 
  radiation from initial (ISR) and final state (FSR) electrons. The accuracy 
  of the implementation of QED radiative
  corrections is checked with the program RADCOR~\cite{radcor},
  using the channel $\EE \ra \EE \mu ^+ \mu ^-$.

  \par
\begin{figure}[htbp]

   \begin{center}
     \includegraphics[height=0.75\textheight]{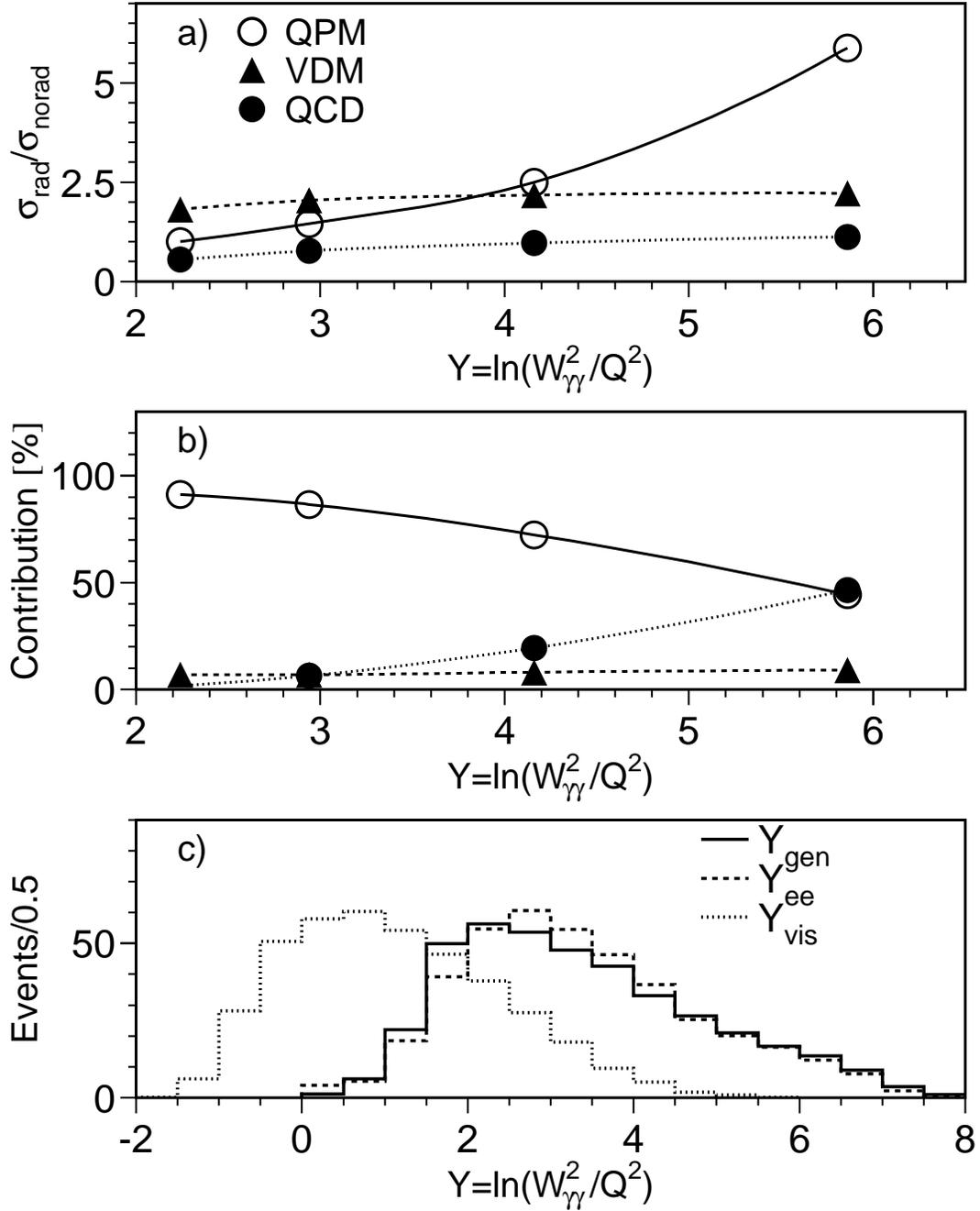}
   \end{center}
 
  \caption[]{a) QED radiative corrections
             as a function of the variable $Y$, 
             for QPM, VDM and QCD processes separately;
             b) the relative contributions of  QPM, VDM and QCD processes
             in the TWOGAM Monte Carlo with QED radiative corrections included and
             c) $Y$ determined using $W_{vis}$ or 
             $W_\mathrm{{ee}}$ compared to the generated value, $Y_{gen}$.
             Lines in a) and b) are drawn to guide the eye.
  \label{fig:rad} }

\newpage
\end{figure}

  The data 
  are mainly sensitive to initial state radiation which modifies the shape of 
  the $Y$ spectrum.
  Since the various processes have different $Y$ dependences, the radiative
  correction affects them differently, as shown in 
  Figure~\ref{fig:rad}a. Here the cross sections are calculated in the
  generator level within the kinematic region defined above.
  The variables $Q^2$ and $W_{\GGG}$ are calculated from the kinematics
  of scattered electrons.
  The relative contributions of QPM, VDM and QCD, as predicted by the 
  TWOGAM program, including QED radiative effects,
  are given in Figure~\ref{fig:rad}b and listed in Table~\ref{tab:process}.
  The VDM contribution is small and almost constant in our kinematical region.
  The resolved photon QCD contribution is negligible at low values of 
  $Y$ and increases to about 50\% at high values.

  \par    
  The dominant backgrounds
  are $\EE \rightarrow \EE \TT$ events, simulated by JAMVG~\cite{verm}, 
  and single-tag two-photon hadronic events, where a hadron mimics
  a scattered electron. 
  Other background processes are simulated 
  by PYTHIA~\cite{pythia} ($\EE \rightarrow   hadrons $),
  KORALZ~\cite{kora} ($\EE~\rightarrow \tau^+\tau^-$) and 
  KORALW~\cite{korw} ($\EE \rightarrow \mathrm{W^+W^-}$).

  All Monte Carlo events are passed through a full detector simulation of the
  L3 detector which uses the GEANT~\cite{GEANT} and
  the GEISHA~\cite{GEISHA} packages and are reconstructed in the same way as 
  the data. Time dependent detector inefficiencies, as monitored during the 
  data taking period, are also simulated.
  The effect of the detector on the generated value of $Y$, $Y_{gen}$, is
  presented in Figure~\ref{fig:rad}c, where the distribution of value
  reconstructed from the hadronic system, $Y_{vis}$, is shown in comparison
  with the quantity $Y_{\mathrm{ee}}$ obtained from scattered electrons.
  The distortion and limited range 
  of the $Y_{vis}$ spectrum, due to the effect of undetected particles, is 
  evident.

\section{Event Selection}
     
    Double-tag two-photon events are recorded by two independent 
    triggers: the central track trigger~\cite{LVT} and the single- and 
    double-tag energy triggers~\cite{LVE} leading to a total trigger
    efficiency greater than 99\%.

\begin{figure}[htbp]

\vfil
   \begin{center}
     \includegraphics[height=0.75\textheight]{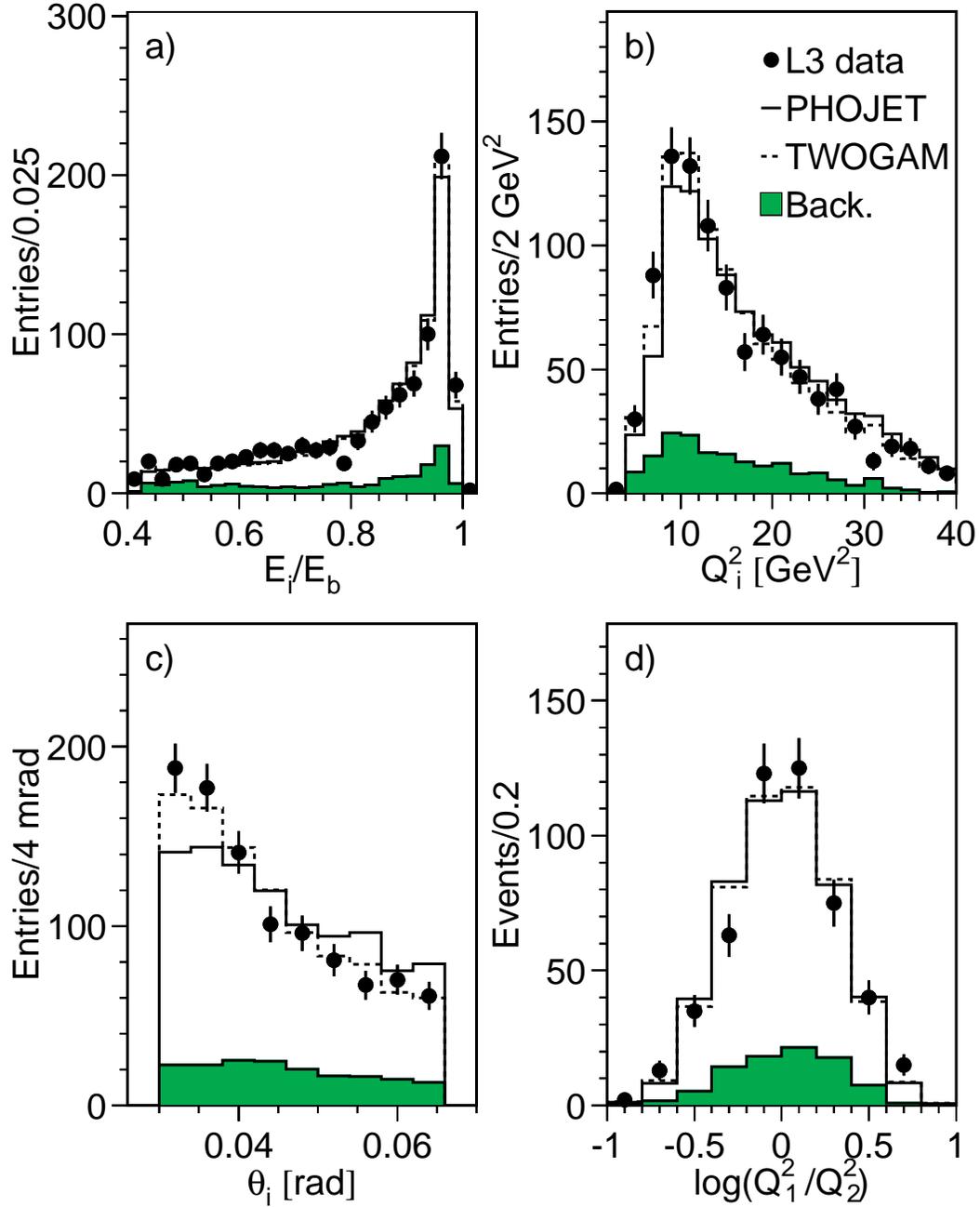}
   \end{center}
 
  \caption[]{Distributions of a) $E_i/E_{b}$, b) $Q^2_i$~, c) $\theta_{i}$ and d) 
             $\log (Q_1^{2}/Q_2^{2})$ for 
             scattered electrons. The data are compared to Monte Carlo 
             predictions, normalised to the total number of events in the data.
             The background is mainly due to 
             $\epem \rightarrow \epem \tau^+\tau^-$ and misidentified single-tag two-photon
             hadronic events. 
  \label{fig:q2} }
\end{figure}

\begin{figure}[htbp]
   \begin{center}
     \includegraphics[height=0.75\textheight]{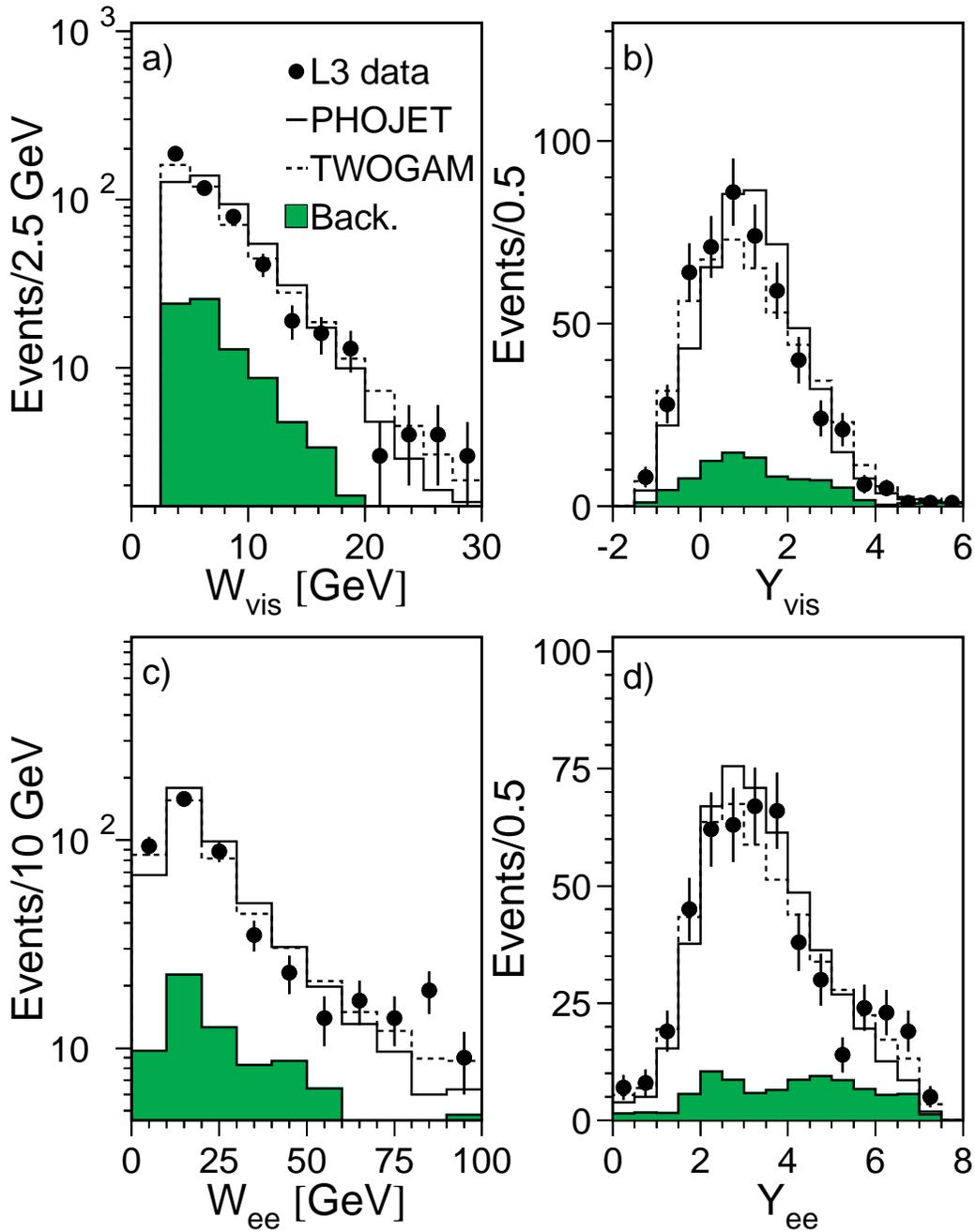}
   \end{center}
   \caption[]{Distributions of 
              a) the effective mass of the detected particles,
                 $W_{vis}$,
              b) $Y_{vis}$,
              c) the missing mass of the two scattered electrons, 
                 $W_{\mathrm{ee}}$ and 
              d) the variable $Y_{\mathrm{ee}}$. The range of $W_{vis}$ and $Y_{vis}$
              is limited to low values due to particles which escape detection.
             The data are compared to Monte Carlo predictions, 
             normalised to the number of data events.
  \label{fig:wtru} }
\vfil
\end{figure}

    Two-photon hadronic event candidates, $\EE \rightarrow \EE   hadrons $,
    are selected using the following criteria:
    \begin{itemize}
     \item There must be two identified electrons, forward and backward, 
           in the small angle electromagnetic calorimeters. Each electron
           is identified as the highest 
           energy cluster in one of the calorimeters, with energy greater 
           than $40 \GeV$. The scattering angles of the two tagged electrons
           have to be in the range 
           $30~\mathrm{mrad} < \theta_{1,2} < 66~\mathrm{mrad}$.
           The opening angle between the momentum vectors of 
           the scattered electrons must be 
           smaller than $179.5^\circ$, to reject Bhabha events. 
           Figure~\ref{fig:q2} shows the distributions of 
           $E_i/E_{b}$, $Q^2_i$, $\theta_i$ and  
           log($Q_{1}^2/Q_{2}^2$) for scattered electrons. 
           TWOGAM describes the shape of the distributions of $\theta_i$ 
           and $Q^2_i$  better than PHOJET.   

     \item The number of particles, defined as tracks and isolated calorimeter
           clusters in the polar angle region $20^\circ<\theta < 160^\circ$, 
           must be greater than 5. The tracks 
           are selected by requiring a transverse momentum greater 
           than $100 \MeV$ and a distance of closest approach, in the 
           transverse plane, to the interaction vertex smaller 
           than 10~mm. Isolated energy clusters are required to have energy 
           greater than $100 \MeV$ and no nearby charged track inside 
           a cone of 35~mrad half-opening angle. 
            
     \item The visible hadronic mass $W_{vis}$, calculated from 
           the four-vectors of all measured particles, must be greater 
           than $2.5 \GeV$ in order to exclude beam-gas and off-momentum
           electron backgrounds. The distributions of $W_{vis}$ and
           of the corresponding variable 
           $Y_{vis}=\ln({W_{vis}^2/\sqrt{Q^2})}$ 
           are compared to Monte Carlo distributions in Figure~\ref{fig:wtru}a 
           and b.        
    \end{itemize}    
    
    After these requirements, 491 events are selected with an estimated 
    background of 49 misidentified single-tag events and 32 events from the 
    process $\EE \rightarrow \EE \tau^+\tau^-$. Other background 
    processes are estimated to contribute 6 events. 
    The variable $W_{\gamma\gamma}$ and
    the corresponding value of Y are calculated from
    the scattered electron variables, $W_{\mathrm{ee}}$ and
    $Y_{\mathrm{ee}}$, shown in Figure~\ref{fig:wtru}c and d.  
    Good agreement is observed with both Monte Carlo generators. 

  
\section{Results}

    The differential cross sections of the $\EE \rightarrow \EE hadrons$
    process with respect to the variables $Q^2$,
    $W_{\GGG}$ and
    $Y$ are measured in the kinematic region:
    \begin{itemize}
     \item $E_{1,2} > 40 \GeV$ and $30~\mathrm{mrad} < \theta_{1,2} < 
           66~\mathrm{mrad}$
     \item $W_{\gamma\gamma} > 5 \GeV $
    \end{itemize}
    The ranges $10 \GeV^2 \le Q^2 \le 32 \GeV^2$, 
    $5 \GeV \le W_{\GGG} \le 100 \GeV$ and $2 \le Y \le 7$ are independently
    investigated. The cross sections are derived in each interval as:
    \begin{equation}
      \Delta \sigma = \frac{\Delta N}{\mathcal{L} \varepsilon}  \xi
    \end{equation}
    where $\Delta N$ is the background subtracted number of events,
    $\mathcal{L}$ is
    the total integrated luminosity and $\varepsilon$ is the selection
    efficiency. This is the ratio of the selected
    number of Monte Carlo events after the full detector simulation to the
    generated number of Monte Carlo events, including QED radiative 
    corrections. An additional multiplicative factor $\xi$, discussed
    above and presented in Figure~\ref{fig:rad}a, corrects the effect of
    QED radiative corrections. The results with and without this correction
    are given in Table~\ref{tab:cro} for different bins together
    with the number of observed events and the selection efficiencies. 
    The size of QED radiative corrections is estimated by TWOGAM using
    the relative proportions of the three components after adjusting the QCD
    component to the data.

   \par
    The systematic uncertainty on the cross sections due to the selection is 
    5\%. It is
    dominated by the effect of a variation of the multiplicity cut from 4 to 6
    particles. The uncertainty from the background estimation of 
    single-tag events is 3.5\% and that due to
    Monte Carlo statistics amounts to 1\%. 
    The uncertainty due to Monte Carlo modelling is estimated 
    as 6.4\% by comparing PHOJET and TWOGAM without QED radiative corrections.
    To check the implementation of QED radiative corrections, the TWOGAM
    predictions for the $\epem \rightarrow\epem \mu^+\mu^-$ process 
    are compared to those of RADCOR. The difference is 
    within 3\% which is included as a systematic uncertainty. 
    The different systematic 
    uncertainties are summarised in Table~\ref{tab:err}.
    The different contribution from QPM, VDM and QCD as function of $Y$
    and $W_{\GGG}$ gives an additional systematic uncertainty. A 20\%
    variation of the QCD component results into an uncertainty of 0.3\%
    at low values of $Y$ and $W_{\GGG}$ and of 5.7\% at large values.
    This uncertainty is about 0.5\% over the full $Q^2$ region.

    \par
\begin{figure}[htbp]
  \begin{center}
  \includegraphics[height=0.75\textheight]{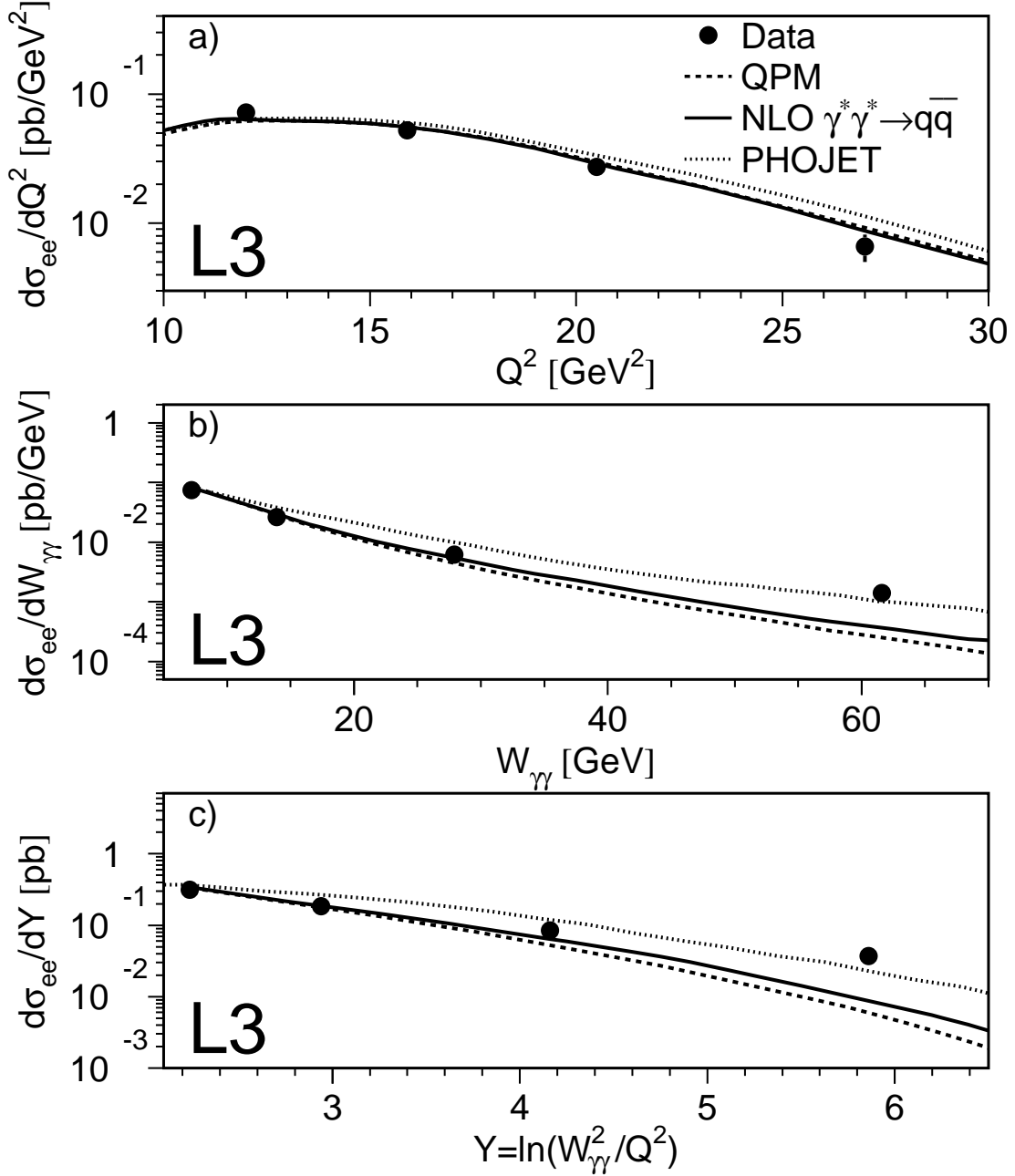}
   \caption[]{The differential cross sections of the  
              $\epem \rightarrow \epem hadrons $ process, in the kinematical region
              defined in the text, after applying QED radiative corrections,
              as a function of
              a) $Q^2$, b) $W_{\GGG}$  and c) $Y$.
              The LO and NLO predictions~\cite{cacciari} for the process 
              $\GG \rightarrow \mathrm{q}\qbar$ are displayed as the dashed and solid
              lines respectively. The dotted line shows the prediction of the
              PHOJET Monte Carlo.
  \label{fig:cro} }
  \end{center}
\end{figure}

     
    The $\EE \rightarrow \EE hadrons$ cross sections after the 
    application of QED radiative
    corrections are compared in Figure~\ref{fig:cro} to the PHOJET Monte Carlo
    and to LO and NLO calculations of 
    $\GG \rightarrow \mathrm{q}\qbar$~\cite{cacciari}.
    In these calculations the mass of quarks is set to zero and 
    $\alpha_{em}$ is fixed to the value for on-shell photons.
    The predictions of these models are also
    listed in Table~\ref{tab:modelcro}.
    These calculations describe well the $Q^2$ dependence of the data.
    For the $W_{\GGG}$ and $Y$ distributions, the QPM calculations describe 
    the data
    except in the last bin, where the experimental cross section 
    exceeds the predictions. Such an excess is expected if the resolved 
    photon QCD processes become important at large $Y$, as illustrated in 
    Figure~\ref{fig:rad}b.
    The predictions of PHOJET, which includes the QPM and QCD processes in
    the framework of the DGLAP equation, also describe the data. 
    A similar behaviour may also be obtained
    by considering the ``hard Pomeron'' contribution in the framework of
    BFKL~\cite{kw} theories, while LO BFKL calculations were found to exceed 
    the experimental values by a large factor~\cite{paper_168}.

    \par
\begin{figure}[htbp]
  \begin{center}
  \includegraphics[height=0.75\textheight]{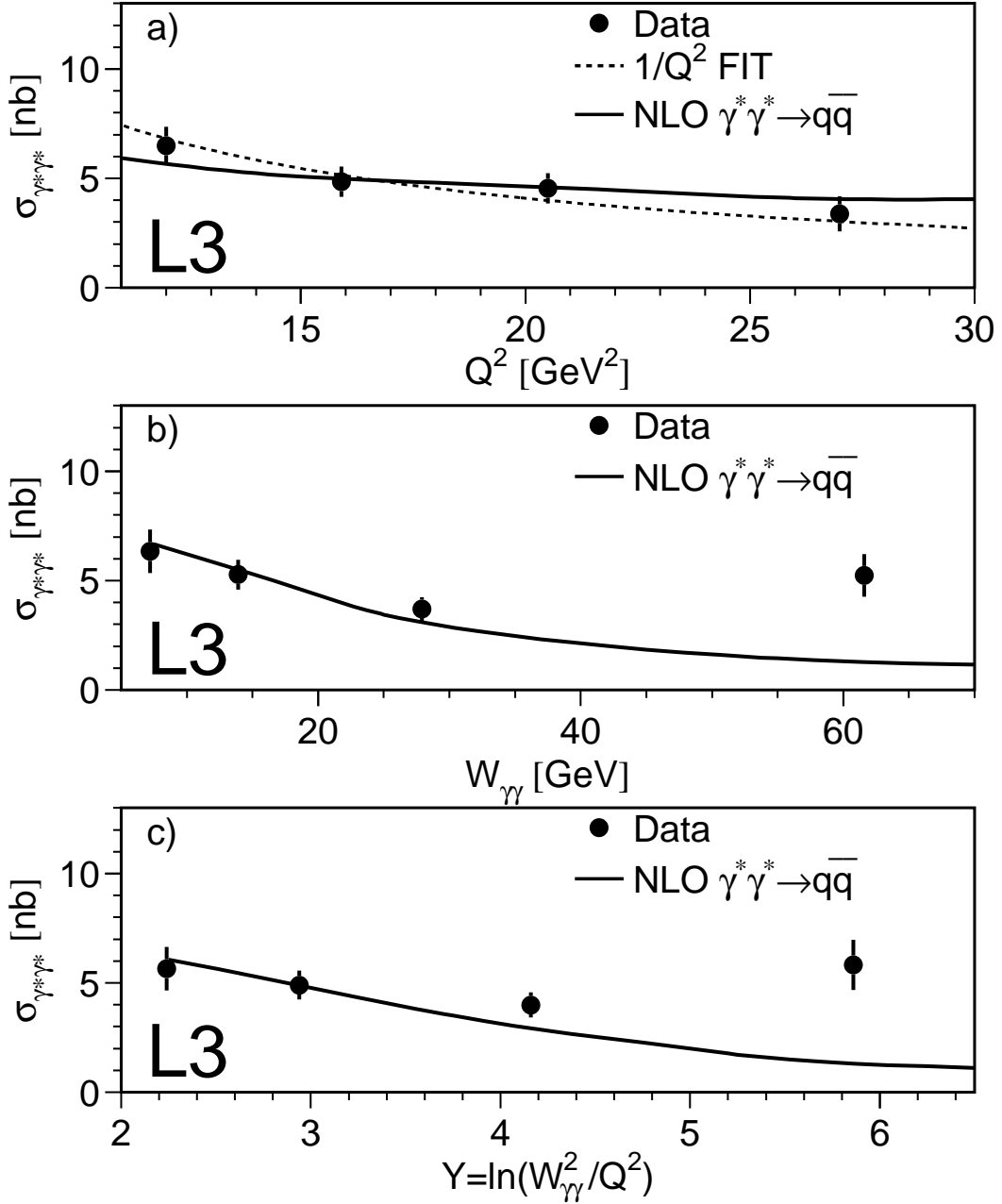}
   \caption[]{Cross sections of the
              $\gamma^*\gamma^* \rightarrow hadrons$ processes as a function 
              of a) $Q^2$, b) $W_{\GGG}$, and c) $Y$  
              in the kinematical region defined in the text, after applying 
              QED radiative corrections.
              The dashed line represents the fit to the data described in the 
              text. The NLO predictions of Reference \citen{cacciari} for the 
              process $\GG \rightarrow  \mathrm{q}\qbar$ are displayed as a 
              solid line.  
  \label{fig:ggcro} }
\end{center}
\end{figure}

        
    From the measurement of the $\epem \rightarrow \epem hadrons$ 
    cross section, $\sigma_{\mathrm{ee}}$, we extract the 
    two-photon cross section, $\sigma_{\gamma^*\gamma^*}$,
    by using the transverse photon luminosity function, $L_{TT}$
    ~\cite{budnev,lumi},
    $\sigma_{\mathrm{ee}} = L_{TT} \sigma_{\GG}$.
    $\sigma_{\GG}$ represents an effective cross section containing 
    contributions from transverse~($T$) and longitudinal~($L$)
    photon polarisations:
    \begin{equation}
     \sigma_{\GG} =  \sigma_{TT} + \epsilon_1\sigma_{LT} +
     \epsilon_2\sigma_{TL} + \epsilon_1 \epsilon_2 \sigma_{LL} +
     \frac{1}{2} \zeta_1\zeta_2 \int{\tau_{TT} \cos 2\tilde{\varphi}
     \mathrm{d} \tilde{\varphi}} -
     4\eta_1\eta_2 \int{\tau_{TS} \cos \tilde{\varphi} 
     \mathrm{d}\tilde{\varphi }}
    \end{equation}
    with
    \begin{equation}
     \zeta_i \sim \eta_i \sim \epsilon_i = 
     \frac { 2 (1-y_i)} {1+(1-y_i)^2}~~,~~~~~ \mathrm{when}~~y_i\muchless 1
    \end{equation}
    where $\tilde{\varphi}$ is the angle between the $\epem$ scattering planes
    in the two-photon centre-of-mass system.
    Using the GALUGA
    Monte Carlo program~\cite{lumi}, the contribution of the 
    interference terms
    $\tau_{TT}$ and $\tau_{TS}$ is found to be negligible 
    for the QPM contribution, when $Y > 3$.
    In the kinematical region studied, the average value of $\epsilon_i$
    is about 0.95.
    The experimental values of $\sigma_{\gamma^*\gamma^*}$ are presented in 
    Table~\ref{tab:crogg} and Figure~\ref{fig:ggcro} in the same ranges 
    considered above with and without QED radiative corrections.
    The measurements as a function of $Q^2$ are fitted by the form 
    $f = A/Q^2$, expected by perturbative QCD~\cite{donn,gg2}. 
    The fit reproduces the
    data well with $A = 81.8 \pm 6.4~\mathrm{nb}/\mathrm{GeV}^2$ and 
    $\chi ^2 / d.o.f = 1.2/3$.
    The average value of $\sigma_{\GG}$ in the
    kinematical region considered is $4.7\pm0.4$~nb. The NLO 
    calculations~\cite{cacciari} predict a  decrease of $\sigma_{\GG}$ as a 
    function of $W_{\GGG}$ or $Y$, which is inconsistent with the measurements
    at large values of $W_{\GGG}$ and $Y$.

\section*{Acknowledgements}
    We thank M.~Przybycien and R.~Nisius for pointing out the importance of
    QED radiative corrections in this process and S.~Todorova for numerous
    discussions about their implementations in the TWOGAM Monte Carlo.

%

\newpage            

 \bibliographystyle{l3stylem}

%
\newpage
\typeout{   }     
\typeout{Using author list for paper 244 -- 247 }
\typeout{$Modified: Jul 31 2001 by smele $}
\typeout{!!!!  This should only be used with document option a4p!!!!}
\typeout{   }
%
%
%
%
%
%

\newcount\tutecount  \tutecount=0
\def\tutenum#1{\global\advance\tutecount by 1 \xdef#1{\the\tutecount}}
\def\tute#1{$^{#1}$}
\tutenum\aachen            
\tutenum\nikhef            
\tutenum\mich              
\tutenum\lapp              
\tutenum\basel             
\tutenum\lsu               
\tutenum\beijing           
\tutenum\berlin            
\tutenum\bologna           
\tutenum\tata              
\tutenum\ne                
\tutenum\bucharest         
\tutenum\budapest          
\tutenum\mit               
\tutenum\panjab            
\tutenum\debrecen          
\tutenum\florence          
\tutenum\cern              
\tutenum\wl                
\tutenum\geneva            
\tutenum\hefei             
\tutenum\lausanne          
\tutenum\lyon              
\tutenum\madrid            
\tutenum\florida           
\tutenum\milan             
\tutenum\moscow            
\tutenum\naples            
\tutenum\cyprus            
\tutenum\nymegen           
\tutenum\caltech           
\tutenum\perugia           
\tutenum\peters            
\tutenum\cmu               
\tutenum\potenza           
\tutenum\prince            
\tutenum\riverside         
\tutenum\rome              
\tutenum\salerno           
\tutenum\ucsd              
\tutenum\sofia             
\tutenum\korea             
\tutenum\utrecht           
\tutenum\purdue            
\tutenum\psinst            
\tutenum\zeuthen           
\tutenum\eth               
\tutenum\hamburg           
\tutenum\taiwan            
\tutenum\tsinghua          

{
\parskip=0pt
\noindent
{\bf The L3 Collaboration:}
\ifx\selectfont\undefined
 \baselineskip=10.8pt
 \baselineskip\baselinestretch\baselineskip
 \normalbaselineskip\baselineskip
 \ixpt
\else
 \fontsize{9}{10.8pt}\selectfont
\fi
\medskip
\tolerance=10000
\hbadness=5000
\raggedright
\hsize=162truemm\hoffset=0mm
\def\r{\rlap,}
\noindent

P.Achard\r\tute\geneva\ 
O.Adriani\r\tute{\florence}\ 
M.Aguilar-Benitez\r\tute\madrid\ 
J.Alcaraz\r\tute{\madrid,\cern}\ 
G.Alemanni\r\tute\lausanne\
J.Allaby\r\tute\cern\
A.Aloisio\r\tute\naples\ 
M.G.Alviggi\r\tute\naples\
H.Anderhub\r\tute\eth\ 
V.P.Andreev\r\tute{\lsu,\peters}\
% ...
\newpage
\begin{list}{A}{\itemsep=0pt plus 0pt minus 0pt\parsep=0pt plus 0pt minus 0pt
                \topsep=0pt plus 0pt minus 0pt}
\item[\aachen]
 I. Physikalisches Institut, RWTH, D-52056 Aachen, FRG$^{\S}$\\
 III. Physikalisches Institut, RWTH, D-52056 Aachen, FRG$^{\S}$
\item[\nikhef] National Institute for High Energy Physics, NIKHEF, 
     and University of Amsterdam, NL-1009 DB Amsterdam, The Netherlands
\item[\mich] University of Michigan, Ann Arbor, MI 48109, USA
\item[\lapp] Laboratoire d'Annecy-le-Vieux de Physique des Particules, 
     LAPP,IN2P3-CNRS, BP 110, F-74941 Annecy-le-Vieux CEDEX, France
\item[\basel] Institute of Physics, University of Basel, CH-4056 Basel,
     Switzerland
\item[\lsu] Louisiana State University, Baton Rouge, LA 70803, USA
\item[\beijing] Institute of High Energy Physics, IHEP, 
  100039 Beijing, China$^{\triangle}$ 
\item[\berlin] Humboldt University, D-10099 Berlin, FRG$^{\S}$
\item[\bologna] University of Bologna and INFN-Sezione di Bologna, 
     I-40126 Bologna, Italy
\item[\tata] Tata Institute of Fundamental Research, Mumbai (Bombay) 400 005, India
\item[\ne] Northeastern University, Boston, MA 02115, USA
\item[\bucharest] Institute of Atomic Physics and University of Bucharest,
     R-76900 Bucharest, Romania
\item[\budapest] Central Research Institute for Physics of the 
     Hungarian Academy of Sciences, H-1525 Budapest 114, Hungary$^{\ddag}$
\item[\mit] Massachusetts Institute of Technology, Cambridge, MA 02139, USA
\item[\panjab] Panjab University, Chandigarh 160 014, India.
\item[\debrecen] KLTE-ATOMKI, H-4010 Debrecen, Hungary$^\P$
\item[\florence] INFN Sezione di Firenze and University of Florence, 
     I-50125 Florence, Italy
\item[\cern] European Laboratory for Particle Physics, CERN, 
     CH-1211 Geneva 23, Switzerland
\item[\wl] World Laboratory, FBLJA  Project, CH-1211 Geneva 23, Switzerland
\item[\geneva] University of Geneva, CH-1211 Geneva 4, Switzerland
\item[\hefei] Chinese University of Science and Technology, USTC,
      Hefei, Anhui 230 029, China$^{\triangle}$
\item[\lausanne] University of Lausanne, CH-1015 Lausanne, Switzerland
\item[\lyon] Institut de Physique Nucl\'eaire de Lyon, 
     IN2P3-CNRS,Universit\'e Claude Bernard, 
     F-69622 Villeurbanne, France
\item[\madrid] Centro de Investigaciones Energ{\'e}ticas, 
     Medioambientales y Tecnol\'ogicas, CIEMAT, E-28040 Madrid,
     Spain${\flat}$ 
\item[\florida] Florida Institute of Technology, Melbourne, FL 32901, USA
\item[\milan] INFN-Sezione di Milano, I-20133 Milan, Italy
\item[\moscow] Institute of Theoretical and Experimental Physics, ITEP, 
     Moscow, Russia
\item[\naples] INFN-Sezione di Napoli and University of Naples, 
     I-80125 Naples, Italy
\item[\cyprus] Department of Physics, University of Cyprus,
     Nicosia, Cyprus
\item[\nymegen] University of Nijmegen and NIKHEF, 
     NL-6525 ED Nijmegen, The Netherlands
\item[\caltech] California Institute of Technology, Pasadena, CA 91125, USA
\item[\perugia] INFN-Sezione di Perugia and Universit\`a Degli 
     Studi di Perugia, I-06100 Perugia, Italy   
\item[\peters] Nuclear Physics Institute, St. Petersburg, Russia
\item[\cmu] Carnegie Mellon University, Pittsburgh, PA 15213, USA
\item[\potenza] INFN-Sezione di Napoli and University of Potenza, 
     I-85100 Potenza, Italy
\item[\prince] Princeton University, Princeton, NJ 08544, USA
\item[\riverside] University of Californa, Riverside, CA 92521, USA
\item[\rome] INFN-Sezione di Roma and University of Rome, ``La Sapienza",
     I-00185 Rome, Italy
\item[\salerno] University and INFN, Salerno, I-84100 Salerno, Italy
\item[\ucsd] University of California, San Diego, CA 92093, USA
\item[\sofia] Bulgarian Academy of Sciences, Central Lab.~of 
     Mechatronics and Instrumentation, BU-1113 Sofia, Bulgaria
\item[\korea]  The Center for High Energy Physics, 
     Kyungpook National University, 702-701 Taegu, Republic of Korea
\item[\utrecht] Utrecht University and NIKHEF, NL-3584 CB Utrecht, 
     The Netherlands
\item[\purdue] Purdue University, West Lafayette, IN 47907, USA
\item[\psinst] Paul Scherrer Institut, PSI, CH-5232 Villigen, Switzerland
\item[\zeuthen] DESY, D-15738 Zeuthen, 
     FRG
\item[\eth] Eidgen\"ossische Technische Hochschule, ETH Z\"urich,
     CH-8093 Z\"urich, Switzerland
\item[\hamburg] University of Hamburg, D-22761 Hamburg, FRG
\item[\taiwan] National Central University, Chung-Li, Taiwan, China
\item[\tsinghua] Department of Physics, National Tsing Hua University,
      Taiwan, China
\item[\S]  Supported by the German Bundesministerium 
        f\"ur Bildung, Wissenschaft, Forschung und Technologie
\item[\ddag] Supported by the Hungarian OTKA fund under contract
numbers T019181, F023259 and T024011.
\item[\P] Also supported by the Hungarian OTKA fund under contract
  number T026178.
\item[$\flat$] Supported also by the Comisi\'on Interministerial de Ciencia y 
        Tecnolog{\'\i}a.
\item[$\sharp$] Also supported by CONICET and Universidad Nacional de La Plata,
        CC 67, 1900 La Plata, Argentina.
\item[$\triangle$] Supported by the National Natural Science
  Foundation of China.
\end{list}
}
\vfill


%
%

\newpage

    \begin{table}[htpb]
      \begin{center} 
        \begin{tabular}{|c|c|c|c|c|}
          \hline
          $\Delta Q^2 (\mathrm{GeV}^2)$ & $10 - 14$ & $14 - 18$ & $18-24$ & $24-32$ \\ \hline
          QPM & 0.778 & 0.844 & 0.890 & 0.919 \\ \hline
          VDM & 0.079 & 0.061 & 0.051 & 0.049 \\ \hline
          QCD & 0.143 & 0.095 & 0.059 & 0.032 \\ \hline \hline
                     
          $\Delta W_{\GGG} (\mathrm{GeV})$ & $5-10$ & $10-20$ & $20-40$ & $40-100$ \\ \hline
          QPM & 0.924 & 0.885 & 0.740 & 0.466 \\ \hline
          VDM & 0.071 & 0.063 & 0.079 & 0.084 \\ \hline
          QCD & 0.005 & 0.052 & 0.181 & 0.450 \\ \hline \hline

          $\Delta Y$ &$2.0-2.5$&$2.5-3.5$&$3.5-5.0$&$5.0-7.0$\\ \hline 
          QPM & 0.913 & 0.866 & 0.724 & 0.443   \\ \hline
          VDM & 0.069 & 0.069 & 0.081 & 0.091   \\ \hline
          QCD & 0.018 & 0.065 & 0.195 & 0.466   \\ \hline         
        \end{tabular}
        \caption[]{Fractional contributions of the three processes, QPM, VDM, and 
                   QCD in different $Q^2$, $W_{\GGG}$ and $Y$ intervals
                   as predicted by the TWOGAM Monte Carlo including
                   QED radiative corrections.
          \label{tab:process} }
      \end{center}
    \end{table}

    \begin{table}[htpb]
      \begin{center}
        \small  
        \begin{tabular}{|c|c|r|c|c|c|c|}
         \hline        
         $\Delta Q ^2$ & $\langle Q^2 \rangle$& & & Before Radiative corrections & After Radiative corrections \\
         $(\mathrm{GeV}^2)$& $(\mathrm{GeV}^2)$ & Events~~~ & $\epsilon$&
         $\mathrm{d\sigma_{\mathrm{ee}} /d}Q ^2 $ (pb/GeV$^2$) &
         $\mathrm{d\sigma_{\mathrm{ee}} /d}Q ^2 $ (pb/GeV$^2$)\\ \hline
         $10-14$& 12.0 & 128.5$\pm$12.4 & 0.58 & $0.0898 \pm 0.0087 \pm 0.0081$ & $0.0718 \pm 0.0070 \pm 0.0061 \pm 0.0022$    \\ \hline
         $14-18$& 15.9 & 102.0$\pm$11.2 & 0.68 & $0.0612 \pm 0.0067 \pm 0.0055$ & $0.0522 \pm 0.0057 \pm 0.0044 \pm 0.0016$    \\ \hline
         $18-24$& 20.5 &  81.3$\pm$\hspace{2.3mm}9.8 & 0.74 & $0.0298 \pm 0.0036 \pm 0.0027$ & $0.0273 \pm 0.0033 \pm 0.0023 \pm 0.0008$    \\ \hline
         $24-32$& 27.0 &  24.8$\pm$\hspace{2.3mm}5.5& 0.77 & $0.0065 \pm 0.0014 \pm 0.0006$ & $0.0066 \pm 0.0014 \pm 0.0006 \pm 0.0002$    \\ \hline \hline
  
         $\Delta W_{\GGG}$ & $\langle W_{\GGG} \rangle$& & & Before Radiative corrections & After Radiative corrections \\
         (GeV) & (GeV) & Events~~~ & $\epsilon$ & $\mathrm{d\sigma_{\mathrm{ee}}
         /d}W_{\GGG}$ (pb/GeV) & $\mathrm{d\sigma_{\mathrm{ee}} /d}W_{\GGG}$ (pb/GeV)\\ \hline
         $\hspace{2mm}5-\hspace{2mm}10$  & ~7.2 & 67.3$\pm$\hspace{2.3mm}8.7 & 0.37 & $0.0594 \pm 0.0076 \pm 0.0053$ & $0.0747 \pm 0.0096 \pm 0.0063 \pm 0.0023$   \\ \hline
         $10-\hspace{2mm}20$  & 13.9 & 135.4$\pm$12.6 & 0.66 & $0.0332 \pm 0.0031 \pm 0.0030$ & $0.0263 \pm 0.0024 \pm 0.0022 \pm 0.0008$   \\ \hline
         $20-\hspace{2mm}40$  & 27.9 & 102.1$\pm$11.1 & 0.72 & $0.0114 \pm 0.0012 \pm 0.0010$ & $0.0062 \pm 0.0007 \pm 0.0005 \pm 0.0003$   \\ \hline
         $40-100$& 61.6 & 65.1$\pm$\hspace{2.3mm}9.8 & 0.67 & $0.0026 \pm 0.0004 \pm 0.0002$ & $0.0014 \pm 0.0002 \pm 0.0001 \pm 0.0001$   \\ \hline \hline
           
         $\Delta Y$ & $\langle Y \rangle$ & & & Before Radiative corrections & After Radiative corrections \\
         & & Events~~~ & $\epsilon$ &$\mathrm{d\sigma_{\mathrm{ee}}/d}Y$ (pb) & 
         $\mathrm{d\sigma_{\mathrm{ee}} /d}Y$ (pb)\\ \hline
         $2.0-2.5$ & 2.2 &  51.6$\pm$\hspace{2.3mm}7.9 & 0.52 & $0.322 \pm 0.049 \pm 0.029$ & $0.315 \pm 0.048 \pm 0.027 \pm 0.009$   \\ \hline
         $2.5-3.5$ & 2.9 & 115.6$\pm$11.4 & 0.73 & $0.258 \pm 0.025 \pm 0.023$ & $0.184 \pm 0.018 \pm 0.016 \pm 0.006$   \\ \hline
         $3.5-5.0$ & 4.2 & 109.4$\pm$11.6 & 0.74 & $0.160 \pm 0.017 \pm 0.014$ & $0.085 \pm 0.009 \pm 0.007 \pm 0.004$   \\ \hline
         $5.0-7.0$ & 5.9 &  53.7$\pm$\hspace{2.3mm}8.9 & 0.63 & $0.069 \pm 0.011 \pm 0.006$ & $0.037 \pm 0.006 \pm 0.003 \pm 0.002$   \\ \hline     
        \end{tabular}
        \normalsize
        \caption[]{Number of events, selection efficiencies, $\varepsilon$, and differential cross sections
                   d$\sigma (\epem \rightarrow \epem   hadrons )$/d$Q^2$,
                   d$\sigma (\epem \rightarrow \epem   hadrons )$/d$W_{\GGG}$ and   
                   d$\sigma (\epem \rightarrow \epem   hadrons )$/d$Y$.
                   All measurements are given before and after applying QED radiative corrections.
                   The first uncertainty is statistical and the second systematic.
                   The third uncertainty represents the effect from QED radiative corrections,
                   including the 3\% from Table~3.                  
       \label{tab:cro} }
      \end{center}
    \end{table}

    \begin{table}[htpb]
      \begin{center} 
        \begin{tabular}{|l|r|}
         \hline           
         Selection procedure      & 5.0\%   \\ 
         Background estimation    & 3.5\% \\
         Monte Carlo statistics   & 1.0\%   \\
         Monte Carlo modelling    & 6.4\% \\
         QED radiative correction & 3.0\%   \\ \hline
        \end{tabular}
        \caption[]{Contributions to the total systematic uncertainties on the measured cross sections.                 
       \label{tab:err} }
      \end{center}
    \end{table}

    \begin{table}[htpb]
      \begin{center} 
        \begin{tabular}{|c|c|c|c|}
         \hline           
         $\Delta Q^2$ & LO $\GG \ra \mathrm{q\bar{q}}$ & NLO $\GG \ra \mathrm{q\bar{q}}$ & PHOJET \\
         $(\mathrm{GeV}^2)$ & 
         $\mathrm{d\sigma_{\mathrm{ee}}/d}Q^2 $ (pb/GeV$^2$) &
         $\mathrm{d\sigma_{\mathrm{ee}}/d}Q^2 $ (pb/GeV$^2$) &
         $\mathrm{d\sigma_{\mathrm{ee}}/d}Q^2 $ (pb/GeV$^2$) \\ \hline
         $10-14$ & 0.0596 & 0.0619 & 0.0623    \\ \hline
         $14-18$ & 0.0547 & 0.0545 & 0.0587    \\ \hline
         $18-24$ & 0.0285 & 0.0279 & 0.0320    \\ \hline
         $24-32$ & 0.0083 & 0.0079 & 0.0100    \\ \hline \hline
  
         $\Delta W_{\GGG}$ & LO $\GG \ra \mathrm{q\bar{q}}$ & NLO $\GG \ra \mathrm{q\bar{q}}$ & PHOJET \\
         (GeV) & 
         $\mathrm{d\sigma_{\mathrm{ee}}/d}W_{\GGG}$ (pb/GeV) & 
         $\mathrm{d\sigma_{\mathrm{ee}}/d}W_{\GGG}$ (pb/GeV) & 
         $\mathrm{d\sigma_{\mathrm{ee}}/d}W_{\GGG}$ (pb/GeV) \\ \hline
         $\hspace{2mm}5-\hspace{2mm}10$  & 0.0831 & 0.0786 & 0.0509   \\ \hline
         $10-\hspace{2mm}20$  & 0.0263 & 0.0269 & 0.0359   \\ \hline
         $20-\hspace{2mm}40$  & 0.0044 & 0.0052 & 0.0094   \\ \hline
         $40-100$& 0.0003 & 0.0004 & 0.0010   \\ \hline \hline
           
         $\Delta Y$ & LO $\GG \ra \mathrm{q\bar{q}}$ & NLO $\GG \ra \mathrm{q\bar{q}}$ & PHOJET \\
         &  
         $\mathrm{d\sigma_{\mathrm{ee}}/d}Y$ (pb) & 
         $\mathrm{d\sigma_{\mathrm{ee}}/d}Y$ (pb) &
         $\mathrm{d\sigma_{\mathrm{ee}}/d}Y$ (pb) \\ \hline
         $2.0-2.5$ & 0.334 & 0.338 & 0.356      \\ \hline
         $2.5-3.5$ & 0.171 & 0.181 & 0.258      \\ \hline
         $3.5-5.0$ & 0.052 & 0.063 & 0.115      \\ \hline
         $5.0-7.0$ & 0.006 & 0.009 & 0.023      \\ \hline
          
        \end{tabular}
        \caption[]{Predictions of LO and NLO $\GG \ra \mathrm{q\bar{q}}$ calculations 
                   and the PHOJET Monte Carlo generator as 
                   a function of $Q^2$, $W_{\GGG}$ and $Y$.                  
       \label{tab:modelcro} }
      \end{center}
    \end{table}

    \begin{table}[htpb]
      \begin{center} 
        \begin{tabular}{|c|c|c|c|c|}
          \hline           
          & & Before Radiative corrections & After Radiative corrections \\
          $\Delta Q^2~(\mathrm{GeV}^2)$ & $\langle Q^2 \rangle~(\mathrm{GeV}^2)$ &
          $\sgg$ (nb) & $\sgg$ (nb)\\ \hline
          $10-14$ & 12.0 & $8.11 \pm 0.79 \pm 0.73$ & $6.49 \pm 0.64 \pm 0.55 \pm 0.20$  \\ \hline
          $14-18$ & 15.9 & $5.68 \pm 0.62 \pm 0.51$ & $4.84 \pm 0.53 \pm 0.41 \pm 0.15$  \\ \hline
          $18-24$ & 20.5 & $4.94 \pm 0.60 \pm 0.45$ & $4.54 \pm 0.55 \pm 0.39 \pm 0.14$  \\ \hline
          $24-32$ & 27.0 & $3.36 \pm 0.74 \pm 0.30$ & $3.38 \pm 0.74 \pm 0.29 \pm 0.10$  \\ \hline
          \hline
          & & Before Radiative corrections & After Radiative corrections  \\
          $\Delta W_{\GGG}$ (GeV) & $\langle W_{\GGG} \rangle$ (GeV) & $\sgg$ (nb) & $\sgg$ (nb)\\ \hline
          $\hspace{2mm}5-\hspace{2mm}10$  & ~7.2 & $5.04 \pm 0.65 \pm 0.45$ & $6.34 \pm 0.82 \pm 0.54 \pm 0.19$  \\ \hline
          $10-\hspace{2mm}20$  & 13.9 & $6.65 \pm 0.62 \pm 0.60$ & $5.27 \pm 0.49 \pm 0.45 \pm 0.16$  \\ \hline
          $20-\hspace{2mm}40$  & 27.9 & $6.84 \pm 0.74 \pm 0.62$ & $3.71 \pm 0.40 \pm 0.32 \pm 0.16$  \\ \hline
          $40-100$& 61.6 & $9.99 \pm 1.50 \pm 0.90$ & $5.24 \pm 0.79 \pm 0.45 \pm 0.34$  \\ \hline
          \hline           
          & & Before Radiative corrections & After Radiative corrections  \\
          $\Delta Y$ & $\langle Y \rangle$ & $\sgg$ (nb) & $\sgg$ (nb)\\ \hline
          $2.0-2.5$ & 2.2 & $5.78 \pm 0.88 \pm 0.52$ & $5.65 \pm 0.86 \pm 0.48 \pm 0.17$ \\ \hline
          $2.5-3.5$ & 2.9 & $6.85 \pm 0.68 \pm 0.62$ & $4.90 \pm 0.48 \pm 0.42 \pm 0.16$ \\ \hline
          $3.5-5.0$ & 4.2 & $7.52 \pm 0.80 \pm 0.68$ & $3.99 \pm 0.42 \pm 0.34 \pm 0.19$ \\ \hline
          $5.0-7.0$ & 5.9 & $10.9 \pm 1.82 \pm 0.98$ & $5.82 \pm 0.97 \pm 0.49 \pm 0.37$ \\ \hline
                    
        \end{tabular}
        \caption[]{The two-photon cross section, $\sigma_{\GG}$, before and
                   after applying QED radiative corrections, as a function of $Q^2$,
                   $W_{\GGG}$ and $Y$.
                   The first uncertainty is statistical and the second systematic. 
                   The third uncertainty represents the effect from QED radiative corrections,
                   including the 3\% from Table~3.                                      
          \label{tab:crogg} }
      \end{center}
    \end{table}

%
%

\newpage

\newpage

\newpage

\end{document}